\documentstyle[12pt]{article}
\title{Heavy Tau Neutrino as the Late Decaying Particle
 in the Cold Dark Matter Scenario\\~\vspace*{3.0em}}
\author{Hisashi Kikuchi and Ernest Ma\\
{}~\vspace*{0.2em}\\
{\normalsize \sl Department of Physics}\\
{\normalsize \sl University of California, Riverside}\\
{\normalsize \sl Riverside, CA 92521} \\
{}~\vspace*{2.0em}
}
\date{UCRHEP-T131\\August 1994\\}
\newcommand{\bea}{\begin{eqnarray}}
\newcommand{\eea}{\end{eqnarray}}
\newcommand{\be}{\begin{equation}}
\newcommand{\ee}{\end{equation}}

\begin{document}
\maketitle
\thispagestyle{empty}

\vfil

\begin{abstract}
The tau neutrino with a mass of about 10 MeV
can be the ``late decaying particle''
in the cold dark matter scenario  for the formation
of structure in the Universe.
We show how this may be realized specifically in the recently proposed doublet
Majoron model.
\end{abstract}

\vfil

\pagebreak
\baselineskip=24pt

\newcommand{\Z}{\mbox{Z\(^0\)}}

\newcommand{\photon}{\mbox{\(\gamma\)}}
\newcommand{\nue}{\mbox{\(\nu_{\rm e}\)}}
\newcommand{\numu}{\mbox{\(\nu_\mu\)}}
\newcommand{\nutau}{\mbox{\(\nu_\tau\)}}
\newcommand{\e}{\mbox{e}}
\newcommand{\eplus}{\mbox{e\(^+\)}}
\newcommand{\majoron}{\mbox{\(\varphi_{\rm L}\)}}
\newcommand{\Znote}{\mbox{Z\(^0\)}}
\newcommand{\vL}{\mbox{\(v_{\rm L}\)}}
\newcommand{\T}{{\rm T}}
\newcommand{\Tf}{\mbox{\( T_{\rm f}\)}}

\newcommand{\teqone}{\mbox{\(t_{\rm EQ1}\)}}
\newcommand{\teqtwo}{\mbox{\(t_{\rm EQ2}\)}}
\newcommand{\Teqone}{\mbox{\(T_{\rm EQ1}\)}}
\newcommand{\Teqtwo}{\mbox{\(T_{\rm EQ2}\)}}
\newcommand{\TD}{\mbox{\(T_{\rm D}\)}}
\newcommand{\lambdaeqone}{\mbox{\(\lambda_{\rm EQ1}\)}}

\newcommand{\rhonutau}{\mbox{\(\rho_{\nu_\tau}\)}}
\newcommand{\rhoR}{\mbox{\(\rho_{\rm R}\)}}
\newcommand{\rhoRRp}{\mbox{\(\rho_{\rm R + R'}\)}}

\newcommand{\mnutau}{\mbox{\(m_{\nu_\tau}\)}}
\newcommand{\Yinf}{\mbox{\(Y_\infty\)}}
\newcommand{\gst}{\mbox{\(g_*\)}}
\newcommand{\gsts}{\mbox{\(g_{* S}\)}}
\newcommand{\mPl}{\mbox{\(m_{\rm Pl}\)}}
\newcommand{\vrel}{\mbox{\(v_{\rm rel}\)}}

\newcommand{\Mpc}{{\rm Mpc}}
\newcommand{\MeV}{{\rm MeV}}

Study of the mass and interaction of neutrinos is a long-standing
subject in particle physics and offers one of the key clues to
possible new phenomena beyond  the standard model.
Majoron models have been attracting a lot of interest in this respect.
They provide the neutrinos with Majorana masses and a new type of interaction
not present in the standard model.
The interaction is due to the coupling of the neutrinos and other
matter to the Majoron,
a Nambu-Goldstone boson associated with the violation of the lepton number
symmetry.
This extra interaction was originally used to facilitate the decay
of a massive neutrino which would otherwise be ruled out
due to the cosmological constraint on the  neutrino mass \cite{chi}.
In a different application,
it allows a stable  massive neutrino to be suitable
as  a dark matter candidate \cite{car}.
In this paper, we examine another possibility that
this interaction may bring about;
the \(\tau\) neutrino (\nutau) as a candidate for the late decaying particle
in the cold dark matter (CDM) scenario for the formation of structure in
the Universe \cite{bar,dav,dod}.

The idea of the late decaying particle was proposed to reconcile a setback
of the CDM model in explaining the formation of
large-scale structure in the Universe \cite{bar,dav}.
This setback became more evident by the recent COBE detection of anisotropy
in the temperature of the cosmic background radiation:
The theoretical prediction for an \(\Omega = 1\) inflationary Universe
on the power spectrum of the density fluctuation
gives a larger power than the observation
at small scales \( \lambda \leq 10 h^{-1} \) Mpc (the Hubble constant
\(H_0 = 100 h \) km/s/Mpc )
once it is normalized
at large scales \( \lambda \sim 10^3 h^{-1} \) Mpc using the
COBE detection \cite{smo}.
A remedy is to delay the time
of matter-radiation equality,
at which sub-horizon-sized fluctuations  begin to
grow.
The delay reduces the power on the small scales,
while the resulting larger horizon size means relatively more
large-scale structure than the simple CDM scenario.

A massive particle species that decays into relativistic
particles  can do this trick.
Since the decreasing rate of the relic energy density of massive matter
is smaller than that of radiation, its energy necessarily dominates
the Universe if it  is sufficiently long-lived.
Its subsequent decay  into relativistic particles
gets the Universe into the radiation-dominant era again with more energy,
and delays the time of forthcoming matter-radiation equality.

In the late decaying particle scenario, we have two matter-radiation
equality epochs:  first the relic \nutau\ dominates the Universe and
then the cold dark matter does. They are separated by a
radiation-dominated era after the \nutau-decay.
We distinguish values of cosmological variables at these epochs
with subscripts `\({}_{\rm EQ1}\)' for the former and `\({}_{\rm EQ2}\)'
for the latter,
e.g., the age of the Universe \teqone,
the temperature \Teqone, the horizon size \lambdaeqone, etc.

Recently, Dodelson et al. examined a scenario in which a massive \(\tau\)
neutrino, with its mass in a range \mnutau\ \(\sim \) 1 -- 10 MeV,
may be the late decaying particle \cite{dod}.
(See also Ref.~\cite{kim} for other candidates in particle physics models.)
An important constraint for this scenario is the one  from
primordial nucleosynthesis:
the equivalent number of massless neutrino species \(N_\nu\)
should be less than
3.3 \cite{wal} or 3.04 \cite{ker}
but a \(\tau\) neutrino of this mass range may possibly contribute
more than this bound \cite{kol2}.

This  difficulty is rather easily evaded in
a Majoron model, thanks to the new interaction between
\nutau\ and the Majoron \majoron.
In the standard model,
the relic abundance \Yinf\ of \nutau, the ratio of its number density
to the entropy density (see \cite{kol} for the definition),
is determined by the speed of the annihilation process
via  \Znote\ exchange compared to the Universe expansion \cite{kol,kol1}.
In  Majoron models the process
\be \nutau \, \nutau \rightarrow  \majoron \,\majoron
\label{pr},\ee
also works to decrease the relic abundance.
The  \nutau -- \majoron\ coupling is proportional to
\(\mnutau / \vL\), where \vL\ is a scale for the lepton number
violation. Thus the process (\ref{pr}) can  be still active even after
the \Znote\ exchange process shuts off if
this \mnutau\ to \vL\ ratio is sufficiently large.
Then the  relic density  can be much smaller than
the one in the standard model.
This can make \nutau\ invisible with respect to the dynamical
evolution of the Universe at the time of primordial nucleosynthesis and
avoid the resulting constraint.  In other words,
the lifetime upper bound of 100 seconds estimated for a heavy \nutau\
to be the late decaying particle in  Ref.~\cite{dod}
is no longer a constraint in this case.

The doublet Majoron (DM) model which we have
proposed recently \cite{kik} is very suitable for the late decaying
particle scenario.
An advantage of this model, compared with the singlet Majoron model,
is that it allows us to use
a smaller scale \vL\ for the lepton
number violation and thus a larger \nutau\ -- \majoron\ coupling
without conflicting with the  basic concept of the
seesaw mechanism \cite{yan1}.
We estimate \Teqone\ for various values of \vL\ and \mnutau, and
show that it indeed provides a possibility that \nutau\ can be
the late decaying particle.

Let us first estimate how low the temperature \Teqone\ should be in order to
satisfy the constraint from primordial nucleosynthesis.
The precise definition for \Teqone\ is the temperature at which
the \(\tau\) neutrino  energy
\(\rhonutau = (2\pi^2/45) \mnutau \Yinf\, \gsts\, T^3\)
 becomes the same as the radiation energy
\( \rhoR = (\pi^2/30) \gst\,T^4 \),
\be \Teqone = {60\over 45} {\gsts\over \gst}\mnutau \Yinf,
\label{Teq}\ee
where \gst\ and \gsts\ are the statistical weights of the light degrees of
freedom for the energy and entropy density, respectively.
They are functions of temperature; specific values of  \gst\ and \gsts\
at a given temperature depend on details of the thermal history of the
Universe.
The values we specifically use correspond to the following situation:
\nutau\ was so heavy, \mnutau\ \(>\) a few MeV, that it was non-relativistic
when the \Znote\ exchange process for its annihilation shuts off;
its abundance is frozen at the temperature \Tf, which is
assumed to take place before electrons (\e) and positrons (\eplus) annihilate
in pairs, i.e., \Tf\ \(>\) a few tenths of MeV.
Thus we use  \(\gst = \gsts = 10 \) (a sum of contributions from
photon (\photon), \e, \eplus, \nue, \numu,
and \majoron) for evaluating  \Tf\ and \Yinf.
After the pair annihilation of \e\ and \eplus,
they reduce to \(\gst = 3.17\) and \(\gsts = 3.64\).
If \nutau\ is relativistic at the termination  of the \Znote\ process,
these values change.
Also there may occur a deviation in the temperatures of \photon\ and
\majoron.
We, however,  neglect these subtleties in this paper.
For lighter neutrino, \( \mnutau \sim  \)  1 MeV, this may cause an error;
but the numerical error in the final results of \Teqone\ expected by
this neglect is at most  30 -- 40 \% and does not affect the discussions
we make in this paper.

The temperature at which primordial nucleosynthesis commences is about
1 MeV \cite{kol}. At this temperature,
we need to satisfy  the condition \(\gst < 11.3 \) or 10.82,
which  correspond to
the bounds
\(N_\nu < 3.3 \) \cite{wal} or 3.04 \cite{ker}, respectively.
Since \( \gst = 10 \) at \(T \sim 1\)  MeV,
the \nutau\ fraction in the energy density must satisfy
\be {\rhonutau \over \rhoR} \leq 0.08. \ee
On the other hand,
the fraction in terms of \Teqone\ is given by
\be {\rhonutau \over \rhoR} \simeq {\gsts(1\; \MeV) \over \gst(1\; \MeV) }
{\gst(\Teqone) \over \gsts(\Teqone) } {\Teqone\over  \mbox{1 MeV} }
\sim {\Teqone\over  \mbox{1 MeV} }\ee
as long as \Tf\ is higher than 1 MeV.
Thus a \(\tau\) neutrino with \Teqone\ less than \(10^{-2}\)
MeV is safe in this respect.

To evaluate \Yinf,
we use  formulas given  by Kolb and Turner \cite{kol},
\be \Yinf = { 3.79 (n+1) x_{\rm f}^{n+1} \over
( \gsts /\gst ^{1/2} ) \mPl \mnutau \sigma_0 }, \label{Yinf}
\ee
and
\bea x_{\rm f} & = & \ln [ 0.038 (n+1) (g/\gst^{1/2}) \mPl \mnutau \sigma_0 ]
\nonumber\\
&& - (n+ 1/2) \ln\{\ln [ 0.038 (n+1) (g/\gst^{1/2}) \mPl \mnutau \sigma_0 ]
\},\label{xf} \eea
where \( \mPl = 1.2 \times 10^{19}\) GeV is the Planck mass,
\(g =2 \) for \nutau, \(n\) and \(\sigma_0\) are
read off from the form of the thermally averaged
cross section for the process
\be \langle \sigma \vrel \rangle = \sigma_0 (T/\mnutau)^n.
\label{ts}\ee
The freezing temperature is given by \(\Tf \simeq \mnutau x_{\rm f}^{-1}\).

The evaluation of the cross section is straightforward.
The appropriate interaction terms are given by
\bea {\cal L} & = &
 { \partial^\mu \majoron \over 2 \vL } (\nutau^\dagger\bar\sigma_\mu\nutau)
+ {\left(\partial^\mu \majoron \right)\left(\partial_\mu \majoron\right)
 \over 2 \vL }\left( \cos\alpha\,\phi_- + \sin\alpha\,\phi_+ \right)
\nonumber\\
&&+ {\mnutau\over 2\vL} \left( \cos\alpha\,\phi_- + \sin\alpha\,\phi_+ \right)
\{ \left( \nutau^{\T} i\sigma^2 \nutau \right) -
\left( \nutau^\dagger i\sigma^2 \nutau^* \right) \},
\eea
where \(\phi_\pm\) are neutral scalar fields,
\(\alpha\) and \(\beta\) are mixing angles, which are
the same as those defined in Ref.~\cite{kik}.
The corresponding cross section in the singlet Majoron model has been evaluated
in  Refs.~\cite{cho}.
We calculate the same Feynman diagrams as those in \cite{cho}:
two diagrams with the  \nutau\ -- \majoron\ vertices and
two scalar-exchange diagrams.
The freezing of the abundance mostly takes place when the
initial two \nutau's are non-relativistic and their energy
is much smaller than the mass of \(\phi_\pm\).
In this energy region, the \(\phi_\pm\)-exchange diagrams are
negligible and we obtain the same result as the one in Ref.~\cite{cho},
\be \sigma = {1\over 96\pi} {\mnutau^2\over\vL^4}{ |\vec p\,|\over E }\ee
in terms of the energy \(E\) and the momentum \(\vec p\) of one of the
initial neutrinos in the center of mass frame.
A characteristic  behavior of \(\sigma\), {\it i.e.} it vanishes at vanishing
\(|\vec p\,| \),  comes from a combined
effect of the statistics of identical particles
and the conservation of both
angular momentum and
``CP''; the latter can be defined as a discrete symmetry of
the \nutau\ -- \majoron\ interaction and forbids the s-wave contribution.
The thermally averaged cross section is given by the integration
over the distribution function for the relative velocity
\( \vrel = 2 |\vec p\,| / E \)\footnote{
Note that the initial neutrinos in the process
(\ref{pr}) are identical Majorana particles. Thus
the event rate of the process per unit  comoving volume is
(1/2) of \(\langle \sigma\vrel \rangle\) defined this way.
In the Boltzman equation, which is the basis to derive the formulas
(\ref{Yinf})-(\ref{xf}) \cite{kol},
this factor 1/2 is cancelled by another  factor 2 that represents
the fact that two neutrinos annihilate in the  process. },
\bea
\langle \sigma\vrel \rangle &\equiv & \int
d^3\vrel \left( {\mnutau\over 4\pi T} \right)^{3/2}
 e^{- m_{\nu_\tau} v_{\rm rel}^2 / 4 T }
\sigma \vrel  \nonumber\\
& = & {1\over 32\pi} {\mnutau T \over \vL^4 }.
\eea
This result gives \(\sigma_0 = (1/32\pi) (\mnutau^2/\vL^4)
\) and  \(n=1\) in (\ref{ts}).

We plot the values of \Teqone\ obtained by Eqs.~(\ref{Teq}), (\ref{Yinf}), and
(\ref{xf}) for various values of
\vL\ and \mnutau\ as  contours in Fig.~1.
The thermal history of the Universe we have assumed
based on the DM model is correct for most  of the range
of values of  \mnutau\ and \vL\ shown in Fig.~1.
Obviously \Teqone\ needs to be higher than  \Teqtwo, which is
about 1 eV, in order that \nutau\ plays the role of the late decaying
particle.
Thus the parameter region in Fig.~1 for accommodating the
late decaying particle is
\be {-9}  < \mbox{log}\left(\Teqone\over\mbox{1 GeV}\right) \le -5.  \ee
(Note that  \nutau\ with \Teqone\ less than 1 eV can be a candidate
of dark matter \cite{car}.)

In the DM model, \vL\ smaller than 10 GeV
predicts a light scalar boson \(\phi_-\), which  can be a rare decay product in
\(\Znote \rightarrow \phi_- + \mbox{(a fermion pair)}\),  and
contradicts the known lower mass bound, about 60 GeV,
for the standard Higgs boson
\cite{kik}\footnote{Since \(\phi_-\)--\Znote--\Znote
coupling is proportional to (\( -\cos \beta \sin \alpha + \sin \beta
\cos\alpha ) \), we can fix this problem
by tuning the two mixing parameters in the model,
i.e., \(\alpha \simeq \beta\); but we will not pursue it in this paper.}.
Thus a region \( \vL \ge 10 \)  GeV is left for the \(\tau\) neutrino
to be the late decaying particle.
The temperature \Teqone\ is, then, 1 -- 10 keV.

\newcommand{\taunutau}{\mbox{\(\tau_{\nu_\tau}\)}}

The lifetime of the \(\tau\) neutrino, \taunutau,  should be adjusted
so that it generates an appropriate amount of radiation energy in its decay.
Let us estimate the required lifetime.
We use the sudden-decay approximation  and assume \nutau\ decays all at once
at the  age  \(t_{\rm D} \simeq \taunutau \) and temperate
\TD.
The total radiation energy density after the decay, including
the decay product R\('\), is
\be
\rhoRRp = {\pi^2\over 30} \gst T^4 \left( 1+{\Teqone\over\TD} \right).
\ee
To fit the predicted power spectrum of the density fluctuation
to the observation, this needs to be about 3
times bigger than the radiation energy in the standard prediction
 \cite{dod}. Thus \( \Teqone /\TD \simeq 2 \).
Taking into account the relation
\( T \propto t^{-2/3}\) in the \nutau-dominated
era, we get \( \taunutau \simeq  3\, \teqone.\)
Since the age at \Teqone\ is
\be
\teqone \simeq 2.4 \times 10^{19} \left( {T_0 \over \Teqone} \right)^2\;
\sec,\ee
where \(T_0 = 2.735 K\) is the present (photon) temperature of the Universe,
\taunutau is \(10^4\) -- \(10^6\) seconds.
The lifetime in the DM model is given by \cite{kik}
\be
\taunutau^{-1} = {1\over 64\pi} |R_{ {\nu_\tau} \nu_a}|^2
{\mnutau^3 \over \vL^2 },
\ee
where \(R_{ {\nu_\tau} \nu_a}\) is the flavor changing matrix element between
\nutau\ and lighter neutrinos (\( \nu_a = \nue, \numu\)).
If we take \vL\ \(\simeq\)  20 GeV and \mnutau\ \(\simeq\) 10 MeV to get an
idea of the magnitude of  \(|R|\),
it resides in a range \( |R| \sim 10^{- 9} \) -- \(10^{-10}\).
The smallness of these values can be regarded as a result of the
seesaw mechanism:
\(|R|\) can be parametrized as \([(\mnutau / M ) \sin \theta ]\)
with \(M\) the mass scale of the  gauge singlet neutrino
and \(\theta\) a mixing angle;
the values we used for \mnutau\ and \vL\ imply  \(M\sim 10 \) TeV and
\(|R| \sim 10^{-6}\, \theta \).

In a late-decaying-particle scenario for structure formation,
we necessarily have an extra small scale corresponding to
the horizon at \teqone.
Its size \lambdaeqone\ at \Teqone, after being scaled up to the present
taking into account the expansion of the Universe, is
\be \lambdaeqone \simeq 4.8 \times 10^5 \left( {T_0 \over \Teqone} \right)
\;\Mpc.
\ee
Thus \(\lambdaeqone \simeq \) 10 -- 100 kpc for the allowed
parameters in the DM model.
The consequence of the existence of this scale for structures in the Universe
needs to be clarifed by further investigations.
It may be related to the dwarf galaxies \cite{bar}.

In summary, we have shown that the DM model represents an
appropriate particle physics model for realizing a possibility that
the \(\tau\) neutrino, with a mass of about 10 MeV and a lifetime
in the range $10^4$ to $10^6$ seconds, is the late decaying
particle in the CDM scenario for the formation  of structure
in  the Universe.

\newpage
\begin{center} {ACKNOWLEDGEMENT}
\end{center}

This work was supported in part by the U.S. Department of Energy under
Grant No. DE-FG03-94ER40837.

\vspace{0.3in}
\begin{center}
\bf Figure caption
\end{center}
\begin{description}
\item[Fig.~1] Contour plot of \( \log [ \Teqone / 1 {\rm GeV} ]\).
The contours correspond to \( \log [ \Teqone / 1 {\rm GeV} ] =
-5, -6, -7, -8, -9, -10 \)  from top to bottom.
\end{description}

\end{document}